\documentclass[aps,pre,twocolumn,nofootinbib,superscriptaddress,showpacs,showkeys]{revtex4-1}
\usepackage{color}
\usepackage{amssymb}
\usepackage{amsmath}
\usepackage{bm}
\usepackage{graphicx}
\usepackage{epstopdf}
\usepackage{newtxtext,newtxmath}
\usepackage{xfrac}
\usepackage{amsfonts}
\usepackage{CJK}
\usepackage{soul}

\definecolor{dkgreen}{rgb}{0,0.6,0}
\definecolor{gray}{rgb}{0.5,0.5,0.5}
\definecolor{mauve}{rgb}{0.58,0,0.82}

\newcommand{\ppo}{{P^{\rm (PO)}(k;m_f)}}

\newcommand{\psf}{{P^{\rm (SF)}(k;\gamma;m_f)}}
\newcommand{\spo}{{S^{\rm (PO)}_f}}
\newcommand{\ssf}{{S^{\rm (SF)}_f}}

\begin{document}

\begin{CJK}{UTF8}{mj}

\setcounter{page}{1}
\title{Degree distributions  under general node removal: Power-law or Poisson?}
\author{Mi Jin Lee (이미진)}
\affiliation{Department of Applied Physics, Hanyang University, Ansan 15588, Korea}
\author{Jung-Ho Kim (김정호)}
\affiliation{Department of Physics, Korea University, Seoul 02841, Korea}
\author{Kwang-Il Goh (고광일)}
\email{kgoh@korea.ac.kr}
\affiliation{Department of Physics, Korea University, Seoul 02841, Korea}
\author{Sang~Hoon~Lee~(이상훈)}
\email{lshlj82@gnu.ac.kr}
\affiliation{Department of Physics and Research Institute of Natural Science, Gyeongsang National University, Jinju 52828, Korea}
\affiliation{Future Convergence Technology Research Institute, Gyeongsang National University, Jinju 52849, Korea}
\author{Seung-Woo Son (손승우)}
\email{sonswoo@hanyang.ac.kr}
\affiliation{Department of Applied Physics, Hanyang University, Ansan 15588, Korea}
\author{Deok-Sun Lee (이덕선)}
\email{deoksunlee@kias.re.kr}
\affiliation{School of Computational Sciences and Center for AI and Natural Sciences, Korea Institute for Advanced Study, Seoul 02455, Korea}

\begin{abstract}
Perturbations made to networked systems may result in partial structural loss, such as a blackout in a power-grid system.   Investigating the resultant disturbance in network properties is quintessential to understand real networks in action. The removal of nodes is a representative disturbance, but previous studies are seemingly contrasting about its effect on arguably the most fundamental network statistic,  the degree distribution. The key question is about the functional form of the degree distributions that can be altered during node removal or sampling, which is decisive in the remaining subnetwork's static and dynamical properties. In this work, we clarify the situation by utilizing the relative entropies with respect to the reference distributions in the Poisson and power-law form. Introducing general sequential node removal processes with continuously different levels of hub protection to encompass a series of scenarios including random removal and preferred or protective removal of the hub, we classify the altered degree distributions starting from various power-law forms by comparing two relative entropy values. From the extensive investigation in various scenarios based on direct node-removal simulations and by solving the rate equation of degree distributions, we discover in the parameter space two distinct regimes, one where the degree distribution is closer to the power-law reference distribution and the other closer to the Poisson distribution.
\end{abstract}

\date{\today}
\maketitle 

\section{Introduction}
\label{sec:introduction}
Complex systems of interacting elements found ubiquitously in nature and society can be represented as networks~\cite{NewmanBook,BarabasiBook}. Depicting a system in the framework of network science helps us to glean insight about the system. For example, the analysis of the network structure residing in a given system can provide valuable leads to uncovering the fundamental question: how the entities form their interactions. Understanding the network structure enlarges our knowledge to the system such as emergent dynamical patterns depending on given structures~\cite{barrat2008dynamical,Dorogovtsev2008}. Examples include outbreaks of epidemics~\cite{PastorSatorras2001, PastorSatorras2015, PhysRevE.99.032309}, blackouts in power grids~\cite{Dobson2007, Andersson20051922, Crucitti200492}, traffic jam on roads~\cite{Zeng23, Daganzo2011278}, and metabolic functions in organisms~\cite{HJeong2000, Stelling2002, Kreimer6976,Wunderlich2006}. Yet, there is always uncertainty in figuring out genuine structures of empirical networks in the real world,  as we inevitably rely on experiments and observations to access the real-world networks. Moreover, real systems naturally undergo changes in structures, caused by the appearance of obsolete parts, random failure, or intentional malicious attacks.

Among the basic statistical properties of networks, the distribution of degree, the number of neighbors connected to each node, has received by far the most attention since the early days due to its decisive role in virtually all aspects of network systems, including phase transitions~\cite{Dorogovtsev2008}, dynamical processes~\cite{Porter2016} and controllability~\cite{YYLiu2011}. In particular, the notion of scale-free (SF) networks was introduced to denote a class of networks exhibiting the degree distribution in a power-law form, $P(k)\sim k^{-\gamma}$ for large values of the degree $k$~\cite{Barabasi1999}, and the degree exponent $\gamma$ has been shown to make (both literally and figuratively) critical differences in the aforementioned aspects~\cite{BarabasiBook,barrat2008dynamical,Dorogovtsev2008,PastorSatorras2001, PastorSatorras2015, PhysRevE.99.032309}. Despite such a far-reaching impact of the power-law degree distributions, their statistical reliability~\cite{Clauset2009} and their implications~\cite{Stumpf2012} have constantly been under scrutiny.
The most recent and intense debate was about whether the SF networks are common or rare in reality~\cite{Broido2019, Voitalov2019}. It is in fact inseparable from the complication caused by the inherent finiteness of real networks, demanding the application of the renowned concept of finite-size scaling in statistical physics~\cite{Serafino2021},  and uncertainty and incompleteness in measurement and  sampling~\cite{Stumpf2005,Stumpf2005PRE,SHLee2006,SWSon2012}. The key question is whether the functional form of the degree distribution remains a power law  under the partial sampling or loss of network structure, informational or real. In fact, a random or biased sampling  is the reverse process to the random failure of or the intentional attack to nodes and the consequent removal of them~\cite{Albert2000, PhysRevLett.85.5468,PhysRevLett.85.4626,*Cohen2001,Shang2021}, where one can observe only the remaining part of the original network.

Partial loss or sampling of a network can be correlated with the centrality of individual nodes~\cite{gallos2005,Morone2015}. Biological evolution proceeds with natural selection that systematically filters species with lower fitness, which is usually related to node centrality, such as degree, represented by the fitness model of SF networks~\cite{Caldarelli2002}. In social systems, there can be more complex scenarios such as the selective vaccination to prevent epidemics or the intervention to the bankruptcy spreading in a financial system,  for which targeting the most appropriate nodes, referring to their degrees and other network properties, is crucial, as it can make an irreversible structural modification. Understanding the structural properties under selective observation is also useful to determine relevant features in graph-based learning techniques~\cite{Scarselli2009}. 

Given such ubiquity of degree-dependent loss or sampling of networks and the importance of characterizing correctly the form of the degree distribution, here we consider as a representative case study the SF networks under the attack to and consequent removal of nodes depending on degree, and investigate how their degree distributions evolve in such circumstances.  Recent studies~\cite{Tishby2020, Tishby2019} have reported that  SF network models such as Barab{\'a}si-Albert (BA) model~\cite{Barabasi1999} and the configuration model~\cite{Newman2001} find their degree distributions converging  towards the Poisson (PO) distribution as nodes are removed randomly or with a preference to the hub nodes, where the Kullback-Leibler (KL) divergence~\cite{KL1951} is used to measure the distance between degree distributions. In other words, SF networks appear to converge to the Erd{\H o}s-R{\'e}nyi (ER) random graph~\cite{Erdos1959} under random node removal. The study makes an impression that a power-law degree distribution changes to the PO distribution under node removal, which may appear inconsistent with the studies claiming that sampling does not in general change essentially the form of the degree distributions of the SF networks~\cite{Stumpf2005,Stumpf2005PRE,SHLee2006,SWSon2012}. 

\begin{figure}[b]
\includegraphics[width=0.8\columnwidth]{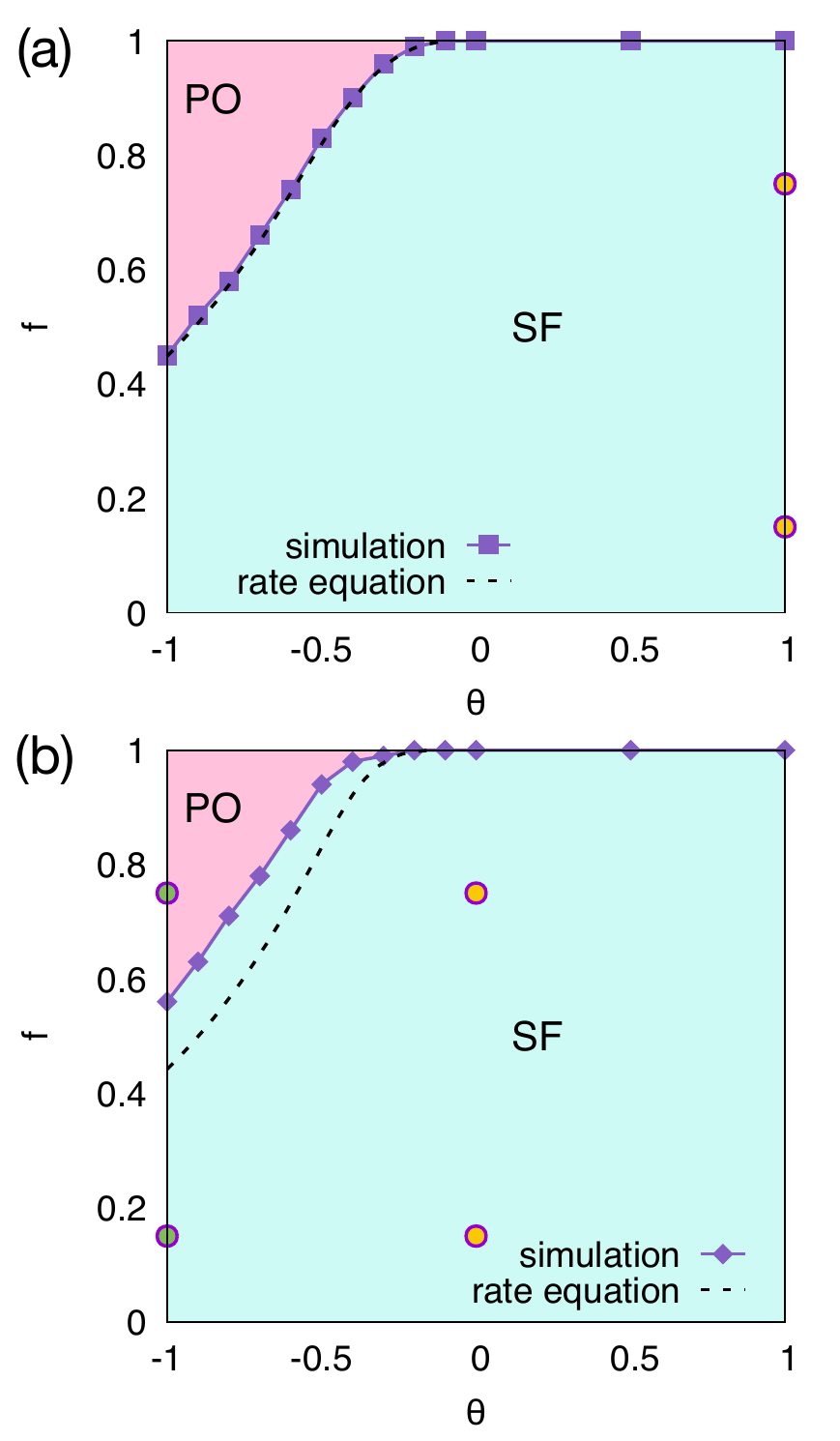}
\caption{Scale-free (SF) and Poisson (PO) regimes on the $f$--$\theta$~plane. The variables $f$ and $\theta$ are the fraction of removed nodes and the parameter that characterizes the level of hub protection in node removal [Eq.~\eqref{eq:removal_prob}], respectively. The degree distribution of the network remaining after node removal is closer to the reference SF (PO) distribution in the SF (PO) regime. The boundary between the two regimes is obtained by comparing the two relative entropies with respect to the two reference distributions  [Eq.~\eqref{eq:entropy}]   from direct simulations (filled points with solid lines dividing the two regimes with distinct colors) and from the rate-equation approach (dashed lines) as detailed in the text. The initial networks are (a) the static-model SF networks~\cite{Goh2001} with the degree exponent $\gamma=2.5$ and (b) the BA  model networks~\cite{Barabasi1999} ($\gamma=3$) both having initial mean degree $10$.  A couple of circular points indicate the parameter sets that are considered in Fig.~\ref{fig:degreedist}.
}
\label{fig:diagram}
\end{figure}

In this work, instead of focusing on a couple of limited cases, we introduce the generalized node removal strategy adjusting continuously  the level of hub protection and study the variation of the degree distribution as we remove nodes for a given hub-protection level. Starting from the SF networks constructed by the static model with the adjustable degree exponent~\cite{Goh2001,Lee2004} and the BA model~\cite{Barabasi1999}, and using the KL divergence, also called \emph{relative entropy},  as a distance measure, we show that the parameter space,  encompassing the limited cases reported in Refs.~\cite{Tishby2019, Tishby2020},  is divided into the regime where the degree distribution  is closer to the PO distribution and the regime where it is closer to the reference SF distribution---the degree distribution of the static model. As shown in Fig.~\ref{fig:diagram},  the SF regime exists, being even larger than the PO one. Therefore, the SF property is preserved over a wide range of parameters. The two reference distributions are shown to be invariant in form and attractive in their respective regimes, under random node removal, suggesting that they can be considered as sort of a stable fixed point in the space of degree distributions. 

The rest of the paper is organized as follows. We introduce our general node-removal scheme in Sec.~\ref{sec:removal}. Applying the removal process for SF networks, we investigate how the degree distribution changes under different types of node removal in the first part of Sec.~\ref{sec:results}. The remaining part of Sec.~\ref{sec:results} is dedicated to presenting the main results by using the relative entropies to systematically explore the distance to the reference distributions, along with the percolation analysis as a representative indicator of modified degree distributions. We conclude the paper with the summary and further discussions in Sec.~\ref{sec:discussion}. In Appendices~\ref{app:static} and \ref{app:equation}, we provide a further numerical analysis and the rate-equation formulation to calculate the degree distribution under node removal, respectively.

\section{General node removal processes} 
\label{sec:removal}

Depending on the scenario that one would like to simulate, the strategy to remove nodes can be differentiated. One can refer to various types of centrality or features of a node for the elimination criterion, e.g., degree and betweenness (centrality)~\cite{Freeman:1977aa,Goh2001}, age and affiliation (feature), etc. In this paper, we take the most intuitive and practical criterion: to remove nodes based on their degree centrality. Despite the simplicity of taking the face values of node centrality, we will demonstrate that the node-removal strategy and the fraction of removed nodes conjointly yield different regimes of the degree distributions of the remaining subnetworks, nicely reconciling the seemingly inconsistent results between Refs.~\cite{Stumpf2005,Stumpf2005PRE,SHLee2006} and Refs.~\cite{Tishby2019, Tishby2020}.

As the degree distributions of many real networks are more heterogeneous than the expected one in completely random graphs~\cite{Erdos1959,NewmanBook,BarabasiBook}, one clear piece of information to determine the nodes to remove can be whether the nodes of interest are rare but highly connected (hubs) or common but sparsely connected. The ``nodes of interest'' can be either the target of removal or the target of protection, depending on the context. In any case, a simple set of degree-dependent node-removal strategies would be setting the node-removal probability as either a monotonically increasing or a monotonically decreasing function of the degree. More complicated scenarios involving non-monotonic behaviors could be considered, but we focus on the monotonic cases throughout our work.
The hub-preferential removal may correspond to a malicious attack towards breaking down a network and also to a vaccination strategy (removing hub susceptible nodes) to prevent or slow the spread of epidemic on social networks~\cite{Hethcote2000, Christakis2010}. In contrast to the hub-preferential removal, there are hub-protecting removal processes, e.g., natural or artificial selection processes that would preferably remove nodes with lower fitness that tend to have small degrees~\cite{Caldarelli2002}. 

To systematically cover possible scenarios, we consider a general rule of removal  with the probability $q(k)$ for a given node of degree $k$ to be removed formulated as
\begin{equation}
q(k) = \frac{(k+1)^{-\theta}}{\sum_{i=1}^{N}(k_i+1)^{-\theta}} = \frac{(k+1)^{-\theta}}{Z} \,,
\label{eq:removal_prob}
\end{equation}
for a network of $N$ nodes, where $\theta$ is a controllable parameter indicating the level of dependence on the degree, and $Z=\sum_{i=1}^{N}(k_i+1)^{-\theta}$ is the normalization factor.  For a well-defined complete set of probability without singularity, even for the isolated nodes ($k=0$), we use $(k+1)$ instead of $k$. With Eq.~\eqref{eq:removal_prob} for a given value of $\theta$, we introduce the node removal process defined as follows. Starting from an original network with $N_0$ nodes and $L_0$ links, one randomly chooses a node $i$ with the probability $q_0(k_i)$ in Eq.~\eqref{eq:removal_prob} with $N_0$ in place of $N$. The selected node and its links are removed from the network, which reduces the degrees of the neighbors by one. The removal probability is updated [$q_{f} (k)={(k+1)^{-\theta} / Z_f}$] reflecting this structural change with $f={1\over N_0}$ indicating the fraction of removed nodes at this stage, with which one of the $(N_0-1)$ remaining  nodes is selected and removed. We repeat these procedures to remove nodes sequentially, increasing  $f$, with the updated removal probability $q_f(k)$, during which we investigate the degree distribution $P_f (k)$ of the remaining subnetwork consisting of  $N_f = (1-f) N_0$ nodes and the $L_f$ links between them with the mean degree $m_f = 2L_f / N_f$ as a function of $f$. 

Our scheme includes the conventional random removal strategy when $\theta = 0$ and  the hub-preferential removal [$q(k) \propto k$] described in Refs.~\cite{Tishby2019,Tishby2020} in the large-degree limit when $\theta = -1$. Positive values of $\theta (>0)$ correspond to the node removal strategy that protects the hub, while one preferentially removes the hubs first for $\theta < 0$. The removal of the node with the largest degree considered in Refs.~\cite{Morone2015,KimJH2020} is achieved in the limit $\theta\to -\infty$, being the extreme case of preferential removal of the hub. In contrary to our processes, the possible reduction of the degrees of the remaining nodes after each node removal is neglected and the removal probability is specified by the initial degrees in some previously studied degree-dependent removal processes~\cite{Albert2000,Cohen2001,gallos2005}.

\begin{figure*}
\includegraphics[width=\textwidth]{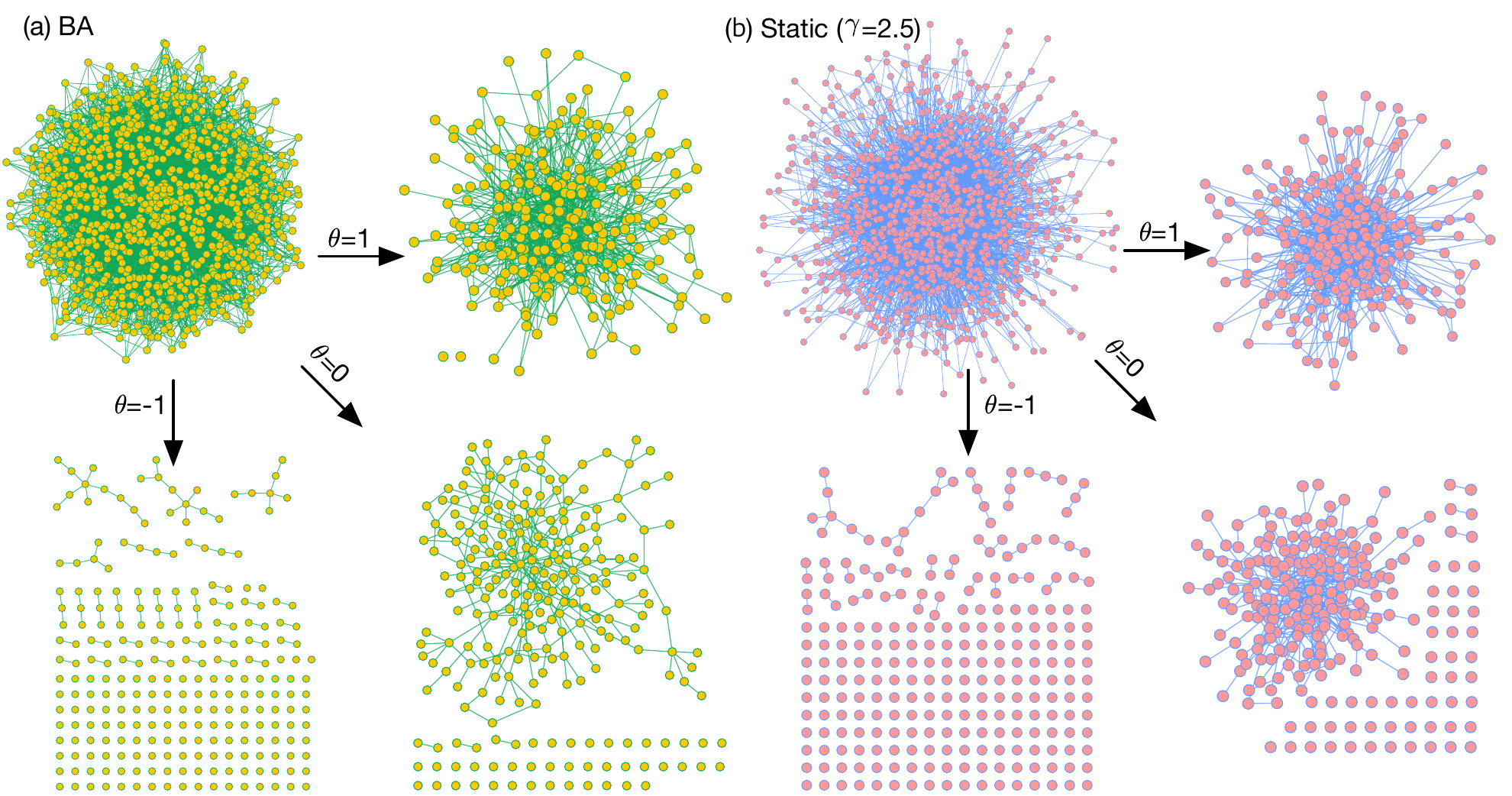}
\caption{The subnetworks remaining after applying the representative cases of node-removal strategy: $\theta=-1$ for the hub-preferential removal, $\theta=0$ for the random removal, and $\theta=1$ for the hub-protecting removal. Panel (a) shows the examples from the original BA network~\cite{Barabasi1999} and panel (b) shows those from the static-model network~\cite{Goh2001} with the degree exponent $\gamma=2.5$. Comparing to each original network ($f=0$) at the top-left corner, the remaining subnetworks contain only $25\%$ of the nodes ($f=0.75$) under the three respective strategies. The size and mean degree of the original networks are $N_{0}=10^3$ and $m_{0}=10$, respectively.}
\label{fig:demo}
\end{figure*}

We show examples of the subnetworks remaining after node removal for the representative cases $\theta=-1, 0,$ and $1$ in Fig.~\ref{fig:demo}. One can find that the original network becomes fragmented into many connected components and isolated nodes as the hubs are preferentially removed ($\theta =-1$), while the original structure seems to be relatively intact for the hub-protecting removal ($\theta =1$). Therefore one can expect different types of degree distributions for the remaining subnetworks with different values of $\theta$. 

Throughout this paper, we focus on the power-law degree distribution $P_0(k)\sim k^{-\gamma}$ with the degree exponent $\gamma$ for the original networks that undergo possible modification during  network shrinkage.
For the original networks, we consider the SF networks constructed by the BA model ($\gamma=3$)~\cite{Barabasi1999} and the static model~\cite{Goh2001} with $\gamma=2.5$ and $3$. The BA model is a hub-preferential growth model with the resulting degree exponent fixed at $\gamma = 3$. On the other hand, the degree exponent is adjustable in the static-model networks~\cite{Goh2001}, which are generated by repeatedly connecting two randomly selected nodes like the ER graphs~\cite{Erdos1959}, however, under the intrinsic inhomogeneous probability of selecting nodes and consequently displaying a power-law degree distribution represented analytically in closed form~\cite{Lee2004}. There is no correlation between the degrees of adjacent nodes in the static-model SF networks except for the inevitable disassortativity for $2<\gamma<3$~\cite{JSLee2006} due to the hub-hub repulsion~\cite{SHYook2005}. Besides the comparison of the node-removal results for the static model with $\gamma=2.5$ and the BA model in the main text, we present in Appendix~\ref{app:static} the results for the static model with $\gamma = 3$ as a comparison to the BA model.  

As the main result, we identify functional forms of modified degree distributions for different node-removal strategies and their consequential percolation properties, by simulating the proposed general node removal processes using the initial size and the mean degree value as $N_{0}=10^5$ and $m_{0}=10$, respectively, for each network. We run the sequential removal process over $100$ realizations for each case, and scrutinize the functional form of the degree distribution $P_f (k)$ during the removal process.  Then, we cross-check the results from the aforementioned node-removal simulations with those based on the rate-equation approach, detailed in Appendix~\ref{app:equation}.

\begin{figure*}
\includegraphics[width=\textwidth]{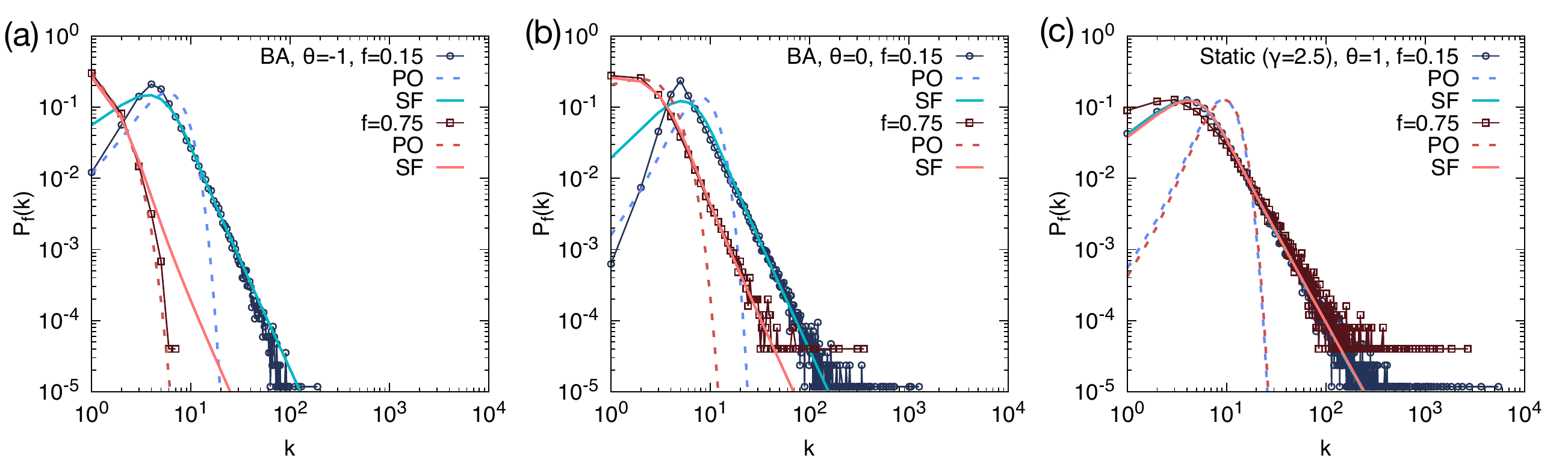}
\caption{Degree distributions of the subnetworks remaining  after node removal.  For representative fractions $f = 0.15$ and $0.75$ of removed nodes, we present the degree distributions (points with lines) of the subnetworks remaining under (a) the hub-preferential removal ($\theta=-1$) and (b) the random removal ($\theta=0$) starting from the BA network~\cite{Barabasi1999} with $\gamma = 3$, and under (c) the hub-protecting removal ($\theta=1$)  from the static-model SF network~\cite{Goh2001} with $\gamma=2.5$.  The dashed and solid lines represent the reference PO and SF distributions in Eqs.~\eqref{eq:poisson} and \eqref{eq:randomsf} with the same mean degrees as the remaining subnetworks. The same degree exponent as the original networks is used for the reference SF distribution. 
}
\label{fig:degreedist}
\end{figure*}

\section{Results}
\label{sec:results}

\subsection{Networks remaining after node removal at different levels of hub protection}
\label{sec:degreedist}

We present the shape of the degree distribution, $P_f (k)$, when the fraction $f$ of the nodes is sequentially removed, for a couple of representative cases in Fig.~\ref{fig:degreedist}.  When the preferential removal of the hub ($\theta=-1$) is applied to the BA networks~\cite{Barabasi1999},  the degree distribution $P_f (k)$ of the remaining subnetwork appears to remain in a power-law form at $f = 0.15$, while the distribution at $f = 0.75$ deviates from the power law, showing a fast decay in the tail part [see Fig.~\ref{fig:degreedist}(a)]. The latter is in accordance with the results in Refs.~\cite{Tishby2019,Tishby2020}. On the other hand, for both the random removal processes ($\theta=0$) and the hub-protecting ($\theta=1$) removal processes, the degree distributions still possess heavy tails even down to $f = 0.75$. See Figs.~\ref{fig:degreedist}(b) and \ref{fig:degreedist}(c). Furthermore, the more heterogeneous degree distributions with $\gamma = 2.5$  seems to conserve the overall shape of the original distribution across the whole range of $k$ under the hub-protecting removal strategy ($\theta = 1$) [see Fig.~\ref{fig:degreedist}(c)]. 

\begin{figure}[b]
\includegraphics[width=0.9\columnwidth]{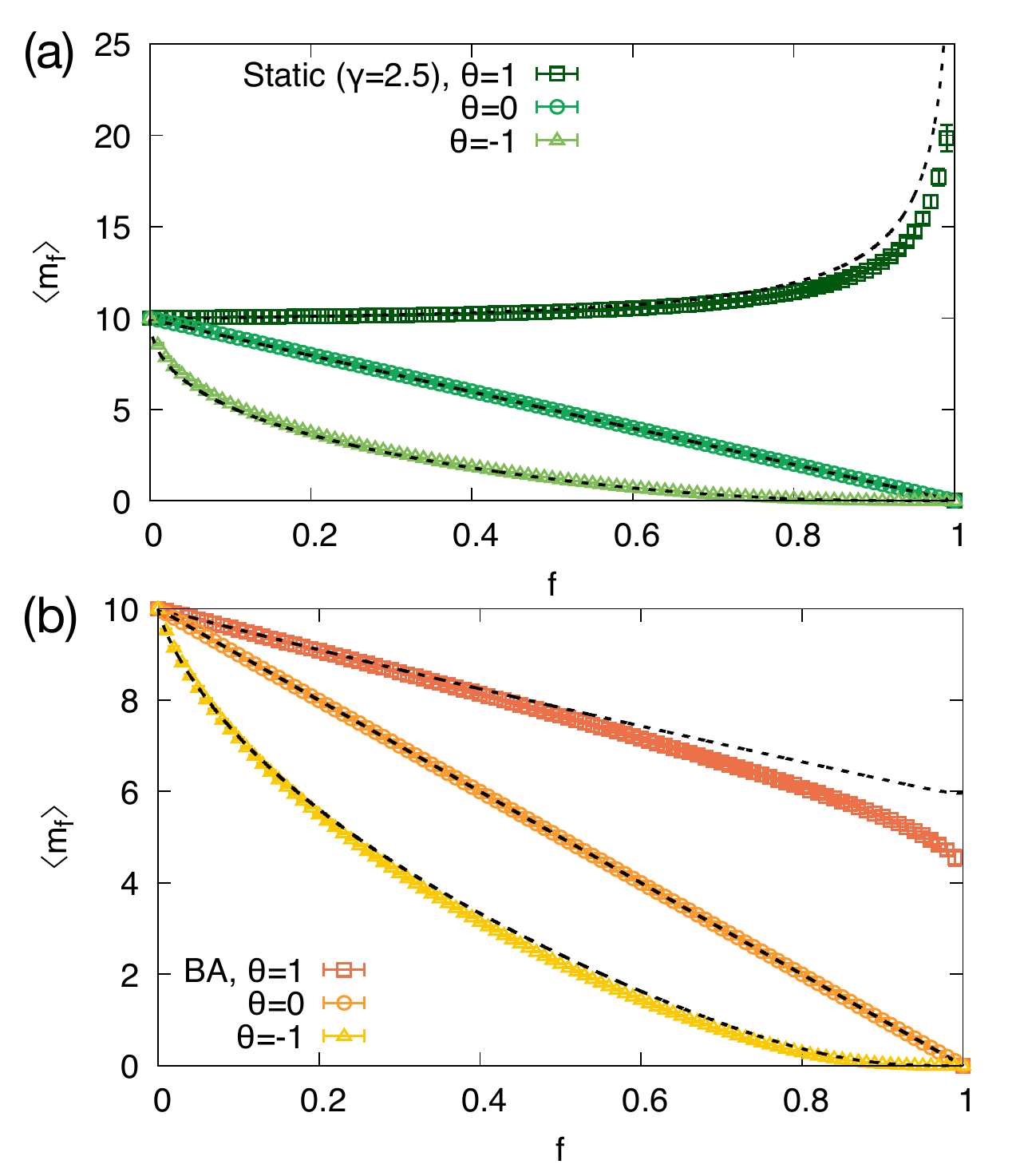}
\caption{Mean degree $\langle m_f \rangle$ versus the fraction $f$ of removed nodes with three representative values of $\theta$ for (a) the static-model SF  networks~\cite{Goh2001} with $\gamma=2.5$ and (b) the BA networks ($\gamma=3.0$)~\cite{Barabasi1999}. Simulation results (points) are obtained by averaging over $100$ realizations, and the error bars, mostly as small as the size of the symbol, indicate the standard deviation. Dashed lines represent the results of the rate-equation approach given in Appendix~\ref{app:equation}.
}
\label{fig:meank}
\end{figure}
  
The mean degree of the remaining subnetworks is plotted as a function of the fraction of removed nodes $f$ for selected values of $\theta$ in Fig.~\ref{fig:meank}.  Apart from $\gamma=2.5$ with $\theta=1$ (protecting hubs), the ensemble-averaged mean degree $\langle m_f\rangle = 2 \langle {L_f / N_f}\rangle$ monotonically decreases as nodes are removed, where $\langle\cdots\rangle$ symbolizes the average over $100$ realizations. The baseline is the random removal with $\theta = 0$, where links are also removed randomly~\cite{SHLee2006} and  the linear relation $\langle m_f \rangle = m_0 (1-f)$ holds as clearly shown in Fig.~\ref{fig:meank}. Deviated from this baseline, we find $\langle m_f \rangle > m_0 (1-f)$ for $\theta > 0$ (hub-protecting removal) and $\langle m_f \rangle < m_0 (1-f)$ for $\theta < 0$ (hub-preferential removal). More specifically, for negative values of $\theta$ (hub-preferential removal), the hubs are removed first, so $\langle m_f \rangle$ decreases rapidly in the early (small-$f$) stage. For the hub-protecting removal ($\theta=1$), $\langle m_f\rangle$ slowly decreases in the early stage or even increases with $f$ in the more heterogeneous network ($\gamma=2.5$) presumably due to the survival of so many hubs.

The goal of the present study is to understand the topology of the remaining network starting from the original SF network displaying $P_0 (k) \sim k^{-\gamma}$. To characterize the functional form of $P_f(k)$ that varies with $f$ and $\theta$, we compare $P_f (k)$ to selected reference distributions as done in Refs.~\cite{Tishby2019,Tishby2020}. In addition to the PO degree distributions of the completely random ER graphs adopted in Refs.~\cite{Tishby2019,Tishby2020}, we also take the degree distribution of the static-model SF networks~\cite{Lee2004}, which we will call the SF distribution, as another reference distribution for comparison. As will be discussed,  both distributions are invariant under the random node removal ($\theta=0$), implying that they are the degree distributions of the {\it random} networks under given constraints, and thus can be featured also by the subnetworks that are expected to be randomized structurally to some extent by successive node removal. We will investigate whether  $P_f(k)$ is closer to the PO distribution or to the SF distribution.  For systematic comparison, we need a similarity or dissimilarity measure between two probability distributions, which will be introduced in the next subsection. Finally, we remark that the static model is used in this work both to construct the original SF networks that will shrink under the node-removal process, and to provide the reference degree distribution (SF distribution). 

\subsection{Reference distributions and relative entropy}
\label{sec:relative_entropy}

The random removal of nodes may wipe out some of structural characteristics of the original networks, and our question can be reduced to whether the heterogeneity of degree characterized by the power-law degree distribution will be  lost or not.  To answer the question, extending the approach introduced in Refs.~\cite{Tishby2019,Tishby2020}, we compare the degree distribution of the remaining subnetwork with two reference distributions, one from the ER graph and the other from the static-model SF networks having the same size and the same mean degree as the remaining subnetwork.  To be specific, the first reference distribution is the PO distribution of the ER random graph~\cite{Erdos1959} represented as 
\begin{equation}
\ppo = \frac{m_f^{k}e^{-m_f}}{k!} \ ,
\label{eq:poisson}
\end{equation}
where $m_f$ is the mean degree of the remaining subnetwork at the removal fraction $f$ and fully determines the functional form of $\ppo$. The second reference distribution is the degree distribution of the static-model SF network represented as~\cite{Lee2004}
\begin{align}
\psf
&=\frac{1}{k!}\frac{d^{k}}{d\omega^k}\tilde{\Gamma}[\gamma, m_f(1-\omega)]\bigg|_{\omega=0} \ ,
\label{eq:randomsf}
\end{align}
where 
$\tilde{\Gamma}[\gamma,m_f(1-\omega)]=\sum_k \psf \omega^k$ is the generating function of $\psf$, and evaluated as 
\begin{equation}
\tilde{\Gamma}(\gamma,x)= (\gamma-1)\left(\frac{\gamma-2}{\gamma-1}x\right)^{\gamma-1}\  \Gamma\left(1-\gamma, \frac{\gamma-2}{\gamma-1}x\right) \nonumber
\end{equation}
with the incomplete Gamma function $\Gamma(s, y)\equiv\int_{y}^{\infty}\,t^{s-1}e^{-t} dt$. The tail part of Eq.~\eqref{eq:randomsf} is approximated as
$\psf\simeq (\gamma-1) \left(m_f {\gamma-2\over \gamma-1}\right)^{\gamma-1} k^{-\gamma}$ 
for large values of $k$, which proves the SF property of the static-model networks~\cite{Lee2004}. The functional form of $\psf$ depends on $m_f$ and the degree exponent $\gamma$. 

We will call the reference distributions in Eqs.~(\ref{eq:poisson}) and (\ref{eq:randomsf}) as the PO and SF distributions, respectively.  Both exhibit a remarkable property; they maintain the functional form under random node removal ($\theta=0$) such that $P_f(k) = \ppo$ if starting from $P_0(k)=P^{\rm (PO)}(k;m_0)$ and similarly $P_f(k) = \psf$ if  $P_0(k)=P^{\rm (SF)}(k;\gamma;m_0)$. A sufficient condition for a degree distribution to display such invariance can be derived as follows, which clarifies why this is the case for the PO and SF distributions. For given $f>0$ and $\theta=0$, it is equally likely that every node has the fraction $f$ of its incident links removed, allowing us to evaluate the degree distribution as $P_f(k) = \sum_{r\geq k} P_0(r) \binom{r}{k} f^{r-k}(1-f)^k$ from the original degree distribution $P_0(k)$~\cite{SHLee2006}, and find the generating function of $P_f(k)$ represented as $g_f(\omega) \equiv \sum_k P_f(k) \omega^k = g_0 (f+(1-f)\omega)$ with $g_0(\omega)=\sum_k P_0(k) \omega^k$.  Let us denote the original degree distribution and its generating function by $P_0 (k;m_0)$ and $g_0(\omega;m_0)$, respectively, to make explicit their dependence on the mean degree $m_0$.  If the generating function depends on $\omega$ and $m_0$ only via $m_0 (1-\omega)$ such that $g_0(\omega;m_0) = \Phi((1-\omega)m_0)$ with a function $\Phi(x)$, then it follows that $g_f(\omega;m_0) = g_0(f+(1-f)\omega;m_0) = \Phi((1-\omega) (1-f)m_0) = \Phi((1-\omega)m_f) = g_0(\omega; m_f)$, where we use the relation $m_f = m_0(1-f)$. This is the case for the PO and SF distributions, which have $\Phi(x) = e^{-x}$ and $\Phi(x) = \tilde{\Gamma} (\gamma,x)$, respectively.

It is of our main concern in the present study whether node removal drives $P_f(k)$ to the reference distributions and if so, to which one of the two $P_f(k)$ gets closer.  
In Refs.~\cite{Tishby2019,Tishby2020}, the convergence of the degree distribution of the original BA networks towards the PO distribution was claimed under node removal  with $\theta=0$ and $\theta=-1$ in our model framework. Considering the SF distribution as well as the PO distribution as the reference will enable us to better understand the convergence.  While we will use the original degree exponent in the reference distribution $\psf$,  the degree distribution $P_f(k)$ of the remaining subnetwork may follow a power law with the degree exponent deviated from the original one as discussed in Ref.~\cite{SHLee2006}, which empirically reported  $P_f(k) \sim k^{-\gamma_f}$ with $\gamma_f$ slightly larger than $\gamma$ in case of $\theta=0$. It is, however, beyond the scope of this paper to accurately measure the altered degree exponent $\gamma_f$, and we use the original exponent $\gamma$ for the reference distribution $\psf$ to focus on the dichotomous distinction of $P_f(k)$ between the PO and the SF distribution. 

For a quantitative measure for the difference between two distributions, we follow the approach of Refs.~\cite{Tishby2019,Tishby2020} and employ the KL divergence~\cite{KL1951} or the \emph{relative entropy}, which represents how a distribution $P(k)$ is different from a reference distribution $P^{\rm (ref)} (k)$. It is defined by $S\left( P \middle\| P^{\rm (ref)} \right) \equiv\sum_{k} P(k) \ln [P(k)/P^{\rm (ref)}(k)]$,  corresponding to  the expected logarithmic difference between $P(k)$ and $P^{\rm (ref)}(k)$ with the weight $P(k)$.  The relative entropy is always non-negative, i.e., $S \left(P \middle\| P^{\rm (ref)} \right) \geq  0$, as known as Gibbs' inequality~\cite{bremaud2012, mackay2003}, and the equality  holds if and only if the two distributions are identical, i.e., $P(k)=P^{\rm (ref)}(k)$ for all $k$ at which $P(k)>0$.  Smaller values of $S$ imply more resemblance between the two distributions. In this work, we compute the following two relative entropies:
\begin{align}
S_f^{\rm (PO)} &\equiv \sum_{k} P_f(k) \ln \left[ \frac{P_f(k)}{\ppo} \right]\, , \nonumber \\
S_f^{\rm (SF)} &\equiv  \sum_{k} P_f(k) \ln \left[ \frac{P_f(k)}{\psf} \right]\, 
\label{eq:entropy}
\end{align}
to measure how similar the node-removed subnetworks' degree distributions are to the one for the ER random graph~\cite{Erdos1959} and for the static-model SF networks~\cite{Goh2001}, respectively. 

As a practical note, the relative entropies in Eq.~\eqref{eq:entropy} do not become singular even if $P_f(k)=0$ for some $k$ because $\lim_{x\to0^{+}}x \ln x=0$ but diverge if $\ppo=0$ or $\psf=0$ while $P_f(k)\neq 0$ for some $k$. Therefore, the reference distributions positive for every possible $k$ are desirable to prevent such singularity. The PO and the SF distribution are positive for all $k\geq 0$, but the BA model~\cite{Barabasi1999,Barabasi1999a} may involve the minimum degree $k_{\min}$ since its degree distribution is defined only for $k\geq k_{\rm min}$. 

\begin{figure*}
\includegraphics[width=\textwidth]{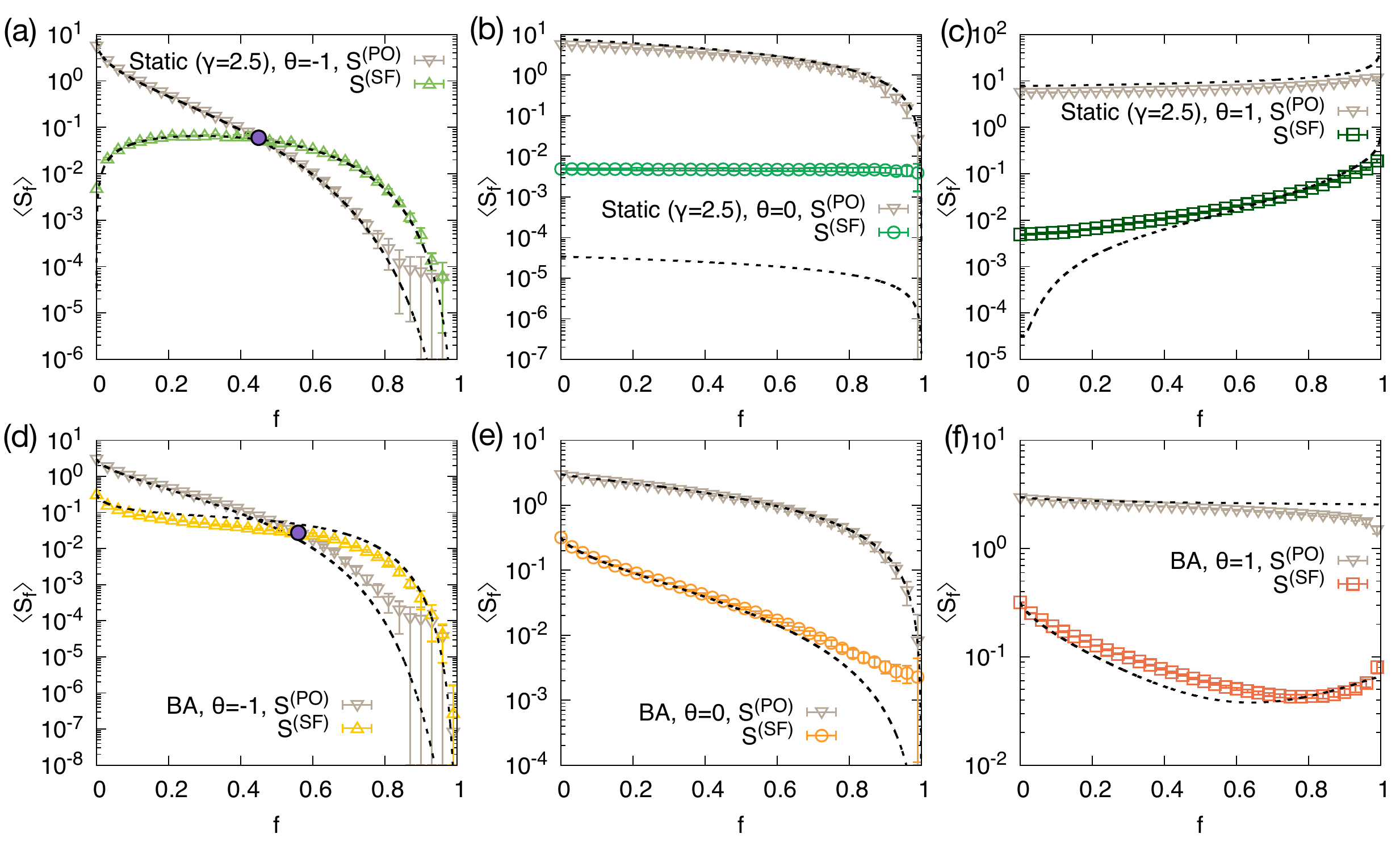}
\caption{Relative entropies $\langle \spo\rangle$ and $\langle\ssf \rangle$ versus the fraction $f$ of removed nodes with [(a)--(c)] the static-model SF networks with $\gamma=2.5$ and [(d)--(f)] the BA networks as the original networks.  Each column stands for a removal process with given $\theta$ as [(a) and (d)] the hub-preferential removal ($\theta=-1$), [(b) and (e)] the random removal ($\theta=0$),  and [(c) and (f)] the hub-protecting removal($\theta=1$). In the hub-preferential removal process, we mark the notable crossing points $f^{*}$ by the purple circles where $\spo=\ssf$ occurs. The points represent the node-removal simulation results, and the dashed lines  represent the rate-equation results. 
}
\label{fig:entropy}
\end{figure*}

\subsection{Classification of degree distributions}
\label{sec:classification}

For initial visual inspection, we overlay the PO distribution $\ppo$ and the SF distribution $\psf$ for given $m_f$ on the empirical data in Fig.~\ref{fig:degreedist}. As discussed in Sec.~\ref{sec:degreedist}, except for the case of hub-preferential removal with $\theta = -1$ up to $f = 0.75$, the empirical distributions $P_f(k)$ look closer to the SF distribution than to the PO distribution. However, the distribution for $\theta = -1$ and $f = 0.75$ illustrated in Fig.~\ref{fig:degreedist}(a) is closer to the PO than to the SF distribution.

For a systematic characterization of the distributions, we present the relative entropies as functions of $f$ in Fig.~\ref{fig:entropy}. Most importantly, $\ssf$ is almost always smaller than the entropy $\spo$ except for the case of the hub-preferential removal with $\theta=-1$ [Figs.~\ref{fig:entropy}(a) and~\ref{fig:entropy}(d)]. For the original BA networks, $\spo$ decreases as $f$ increases for all the considered values of $\theta$,  and therefore one can be impressed that  the degree distribution evolves under node removal towards the PO distribution~\cite{Tishby2019,Tishby2020}. However, the $\ssf$ of the original BA networks also decreases with $f$, except for a slight increase near $f=1$ with $\theta=1$, and moreover, $\ssf$ remains smaller than $\spo$ for $\theta=0$ and $\theta=1$.  For either of the two original SF networks, the comparison of $S^{\rm (PO)}_f$ and $S^{\rm (SF)}_f$ reveals that all the way up to $f \to 1$  with $\theta=0$ or $\theta=1$, the $P_f(k)$ distribution is never closer to $\ppo$ than to $\psf$ [see Figs.~\ref{fig:entropy}(b),~\ref{fig:entropy}(c),~\ref{fig:entropy}(e), and~\ref{fig:entropy}(f)]. 
This argument is stronger than it seems. Although the degree exponent could be effectively changed~\cite{SHLee2006} during the node-removal process as we mentioned earlier, the $P_f(k)$ distribution is found to be closer to $\psf$ with the original exponent $\gamma$ than to $\ppo$, implying that in reality it must be even closer to the SF distribution with a more fine-tuned degree exponent $\gamma_f$. 

Looking more closely at the case of random node removal with $\theta = 0$ shown in Figs.~\ref{fig:entropy}(b) and \ref{fig:entropy}(e), we observe that the behavior of $\ssf$ as a function of $f$ is variable with the original network, but the main conclusion is shared that the degree distributions converge closer to the SF distribution rather than the PO one.  When the original network is the static-model network~\cite{Goh2001}, the degree distribution of the node-removed networks maintains its original functional form, which results in $\ssf$ almost constant against the variation of $f$ at a relatively low level [Fig.~\ref{fig:entropy}(b)] as expected from the discussion on its invariance in Sec.~\ref{sec:relative_entropy}. We remark that $\ssf$ is small but not zero even for $f=0$ as a realization of the finite static-model network may have its  degree distribution deviating slightly from Eq.~(\ref{eq:randomsf}) that is expected to be valid in the $N_0\to\infty$ limit. The original BA networks find $\ssf$ decreasing with $f$ [Fig.~\ref{fig:entropy}(e)] meaning that the random node removal makes the remaining subnetwork resemble the static-model SF networks~\cite{Lee2004}. It is also consistent with previous reporting that the degree distribution of the BA model is relatively stable under random omission of nodes or links such that $\gamma_f \simeq \gamma_{f=0} = 3$~\cite{SHLee2006}. 

Under the hub-protecting removal ($\theta=1$), the original static-model networks find $\ssf$  increasing with $f$ [Fig.~\ref{fig:entropy}(c)] since more hubs  are present than expected in the static model for given mean degree. In addition, more nodes are found to have small degrees than expected  [Fig.~\ref{fig:degreedist}(c)]. We see again that $\ssf$ remains smaller than $\spo$, implying that the SF distribution describes better the degree distribution than the PO one does. While the original BA networks first get closer to and then slightly away from  the SF distribution with the original degree exponent, the degree distribution is always closer to the SF distribution than to the PO distribution [Fig.~\ref{fig:entropy}(f)]. The observation of $\ssf<\spo$ with both original networks for $\theta=1$ is understandable, as the hub-protecting procedure effectively tends to preserve the tail part of the power-law distribution.  Although it is not exactly the same procedure, the snowball sampling reported in Ref.~\cite{SHLee2006} effectively samples the hubs first (in terms of removal that would correspond to removing the hubs later) by consecutively following the neighboring links, and yields more heterogeneous distributions characterized by $\gamma_f < \gamma$. 

Finally, when we remove nodes in a hub-preferential manner, the hubs are eliminated in the early stage, weakening the tail part of the distribution so that for the fraction $f$ large enough, the distribution becomes closer to $\ppo$ than to $\psf$. See Figs.~\ref{fig:entropy}(a) and~\ref{fig:entropy}(d). Based on numerical simulations, we actually pinpoint the crossing point $f^*$ [marked by big filled circles in Figs.~\ref{fig:entropy}(a) and~\ref{fig:entropy}(d)] where the distribution starts to be closer to $\ppo$ than to $\psf$.  The crossing point constitutes a boundary for the classification of degree distributions. We extensively explore the crossover point $f^*(\theta)$ beyond which $\spo$ is smaller than $\ssf$  with each network, and we obtain the regime diagram with respect to the fraction $f$ of removed nodes and the hub-protection level parameter $\theta$ in Fig.~\ref{fig:diagram} with  $f^*(\theta)$ giving the boundary between the PO and the SF regime, where the degree distribution is closer to the PO and the SF distribution, respectively.

The degree distribution with a smaller degree exponent has a fatter tail than that with a larger degree exponent, and thus one might expect the PO regime, located for sufficiently large $f$ and negative $\theta$, to be narrower if the original degree distribution has a smaller degree exponent. Comparing the regime diagram between the original SF networks with $\gamma=2.5$ and $3$, however, we find that the PO regime is larger for $\gamma=2.5$ than for $\gamma=3$ as shown in Fig.~\ref{fig:diagram} and in Appendix~\ref{app:static}.  Such a counterintuitive result originates in the different impacts of a hub-preferential removal for the degree distribution depending on the degree exponent in heterogeneous networks.  

The exact boundary between the PO and the SF regime would rely on the aforementioned altered degree exponent $\gamma_f$ for $\ssf$ and the type of entropy measures (e.g., one can use other measures such as $f$-divergence~\cite{fdiv}). Again, we would like to stress that our main interest is  if the degree distribution of SF networks is modified enough to be identified as completely different types of distributions such as the PO distribution as a result of node removal  with different levels of hub preference. Our conclusion is that a certain (positive) threshold level of hub-preference combined with a certain threshold fraction of removed nodes (summarized in the regime diagram in Fig.~\ref{fig:diagram}) is required for the degree distribution of the node-removed network to be classified as the PO distribution. In the SF regime with the parameters below those thresholds, the degree distributions remain closer to the SF distribution, another invariant reference distribution identified in the present study.

To check the finite-size effect, we have varied the system size $N$ (the number of nodes) and compared the regime diagrams. First, for the values of $\theta$ where both the PO and SF regimes exist, the crossing point $f^{*}$ of the relative-entropy curves becomes smaller and eventually saturated as $N$ increases. In other words, for larger values of $N$, the PO regime becomes larger in the regime diagram, and one can presumably expect the establishment of the boundary curves in the thermodynamic limit $N \to \infty$. To complement these simulation-based results, we take the rate-equation approach~\cite{Tishby2019, Tishby2020, KimJH2020, KimJH2022} and obtain by numerical integration the degree distribution approximated in the thermodynamic limit as shown in Appendix~\ref{app:equation}. Using the result, we compute the mean degrees, the relative entropies, and the regime diagrams, which are shown in Figs.~\ref{fig:diagram}, \ref{fig:meank}, and \ref{fig:entropy}.  As the correlation of the degrees of adjacent nodes is neglected in the rate-equation approach,  the obtained results for the original BA networks show deviations from the simulation results e.g., in Fig.~\ref{fig:diagram}(b). In addition, the degree distributions are obtained up to a finite maximum degree, which brings a non-zero relative entropy with respect to the SF distribution even for the original static-model networks as shown in Fig.~\ref{fig:entropy}(b). See Appendix~\ref{app:equation} for more details. 

Finally, to verify the robustness of our qualitatively different results between the hub-protecting and hub-preferential removal processes ($\theta > 0$ versus $\theta < 0$), we have extended our analysis to selected values of $\theta$ in  the range of $\theta<-1$ and $\theta>1$. The phase boundary $f^*(\theta)$ decreases as $\theta$ decreases beyond $-1$ while it remains at $1$ for $\theta>0$. In the most extreme case of hub-protecting or hub-preferential removal strategies corresponding to  $\theta\to \infty$ or $\theta\to -\infty$, respectively,  one deterministically (except for inevitable stochasticity in the order of removal of multiple nodes with exactly the same degree~\cite{KimJH2020}) removes the node with the smallest or largest degree at each time step, respectively. The case of $\theta \to \infty$ in fact corresponds to the well-known $k$-core decomposition process~\cite{YXKong2019}, where nodes are sequentially removed in ascending order of degree. Our simulation results suggest that $f^*(\theta\to\infty) \to 1$ and $f^*(\theta\to-\infty)$ is close to $0$, implying that the SF (PO) regime dominates such extreme hub-protecting (hub-preferential) removal process. In other words, our main result for the range $\theta \in [-1,1]$ is naturally extended to the most extreme cases of hub-protecting or hub-preferential processes, i.e., the completely SF-dominating regime holds for all values of $\theta > 0$ and the PO regime is increased as the $\theta$ value decreases for $\theta <0$.  

\subsection{Giant component under general node removal processes}
\label{app:gc}

\begin{figure}[b]
\includegraphics[width=\columnwidth]{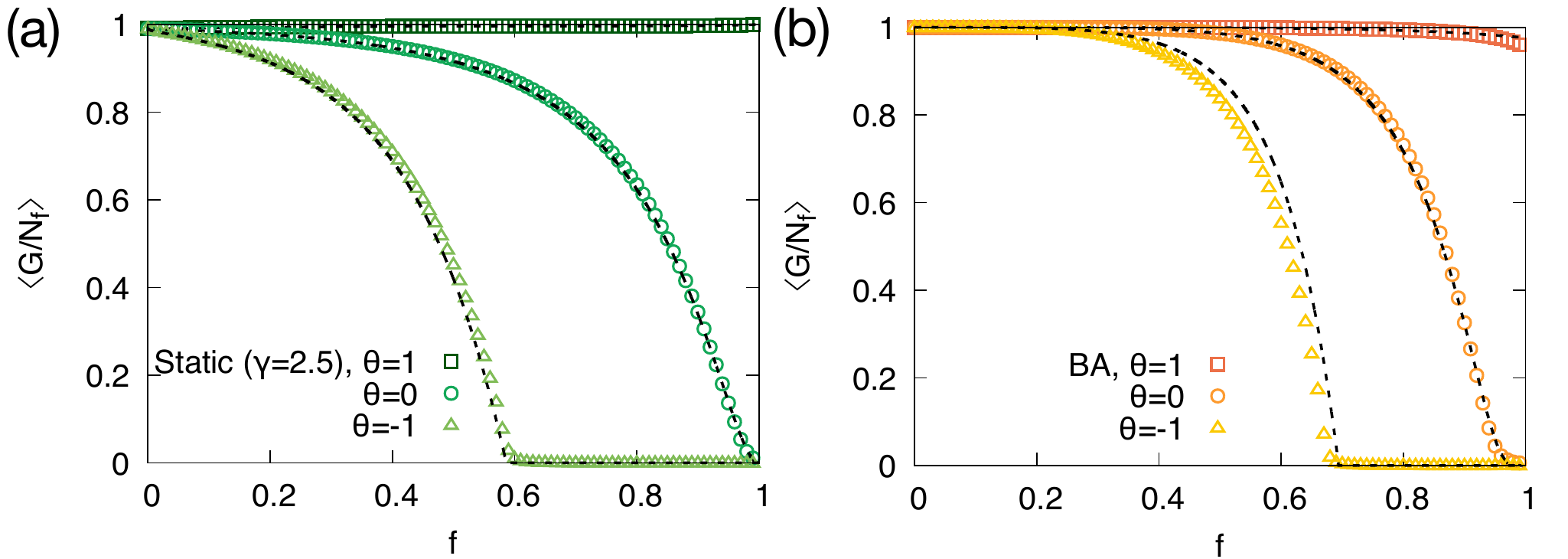}
\caption{Relative size of the GCC, $\langle {G / N_f }\rangle$, versus the fraction $f$ of removed nodes  for (a) the static-model SF networks~\cite{Goh2001} with $\gamma=2.5$ and (b) the BA networks~\cite{Barabasi1999}. Here $G$ is the number of nodes included in the GCC and $N_f = N_0(1-f)$ is the total number of remaining nodes with $N_0$ the original number of nodes. The average and the standard deviation over 100 realizations are represented by the points and the error bars. The dashed lines represent the results obtained by applying the generating function method~\cite{PhysRevLett.85.5468, Newman2001}  to the degree distributions from the rate-equation approach as  detailed in Appendix~\ref{app:equation}.
}
\label{fig:app_gc}
\end{figure}

The existence of a giant connected component (GCC) is a necessary condition for the overall functioning of a networked system. Therefore, the fraction of nodes included in the GCC has been utilized as an elementary measure of how well a system can work while nodes are removed by failure or attack~\cite{Albert2000} and as the central order parameter in the percolation properties of networks~\cite{PhysRevLett.85.5468,PhysRevLett.85.4626,*Cohen2001,gallos2005, Shang2021,Morone2015}.  In particular, the behavior of the relative size of the GCC  is indicative of distinct types of networks; the GCC emerges at the critical mean degree equal to one, $m_c=1$, for the ER random graph~\cite{Erdos1959,PhysRevLett.85.5468} while the percolation threshold $m_c$ remains zero (always percolating as long as $m>0$) for SF networks with $\gamma \leq 3$ and increases from $0$ to $1$ continuously as  $\gamma$ increases from $3$ to infinity~\cite{PhysRevLett.85.4626,*Cohen2001, Lee2004, Dorogovtsev2008}. 

The classification of the type of the degree distribution can help understand the possibly different behaviors of the GCC under attack depending on the level of hub protection parameterized by $\theta$ in  our node-removal scheme. If an original SF network with $\gamma\leq 3$ is changed to the random-graph-like structure at some point of $f$, we can expect that the GCC shrinks before reaching $f\to 1$. That is precisely what happens in Fig.~\ref{fig:app_gc} for $\theta = -1$ in both the original static-model networks with $\gamma=2.5$ and the BA networks with $\gamma=3$. On the contrary, in the range of $\theta$ where the SF regime is preserved throughout $f$, the GCC is well preserved up to $f \to 1$, also clearly shown in Fig.~\ref{fig:app_gc}.  See Fig.~\ref{fig:demo} for the fragmented structures different depending on $\theta$. 

\section{Summary and Discussions}
\label{sec:discussion}

We have explored the degree distribution of the node-removed subnetworks originating from SF networks, under the generalized node-removal processes with different levels of hub preference or protection. By taking the relative entropy as an inspection tool to distinguish between the PO and SF distribution, we have systematically characterized the resultant degree distribution in wide ranges of hub-preference and the fraction of removed nodes. In particular, we have specified the regime where the degree distribution is actually closer to the PO distribution than to the SF distribution, beyond the finding of its asymptotic approaching to the PO distribution~\cite{Tishby2019,Tishby2020}, but only for the hub-preferential removal scheme with $\theta < 0$. For $\theta\ge 0$, we have discovered that the resultant distribution is always closer to the SF distribution than to the PO distribution, which is consistent with the random sampling scheme~\cite{SHLee2006} ($\theta = 0$). Besides the relative entropy to quantify the relative distance between the distributions, we have observed the change of GCC throughout the node-removal processes to verify the modified structures. 

As real networks are always under potential failure and/or intentional attack, characterizing the degree distribution of networks under various types of node-removal processes is crucial as the degree distribution is by far one of the most fundamental structural properties that govern the dynamical and functional aspects of networks. Although our simple scheme cannot capture all of the complicated scenarios in reality, we believe that inspecting a wide range of different levels of hub-preference and the fraction of removed nodes is a first step to proceed.  A natural extension can be the link removal process introduced in Ref.~\cite{config2019}, where the non-random case $\theta \neq 0$ can be realized in various ways to reflect different possible scenarios. In addition, one can try different power-law degree distributions from those considered in this work. As mentioned in Sec.~\ref{sec:relative_entropy}, there are various forms of the degree distributions of SF networks, including the BA model~\cite{Barabasi1999} and the configuration model~\cite{Newman2001}.

Finally, the journey to the breakdown of networks in terms of percolation theory is definitely worth further investigation, as we have a whole toolkit to apply throughout $f$ including the identification of the percolation threshold, the estimation of the critical exponents, and more comprehensive finite-size scaling analysis, etc, the results of which will be reported elsewhere. Beyond the scope of degree distribution, the percolation analysis would wide open the door to characterize the networks with partial loss in general. 

\begin{acknowledgments}
This work was supported by grants from the National Research Foundation of Korea (NRF) funded by the Korean Government (No. NRF-2021R1C1C1007918 (M.J.L.), NRF-2020R1A2C2003669 (K.-I.G.), NRF-2021R1C1C1004132 (S.H.L.), NRF-2020R1A2C2010875 (S.-W.S.), and NRF-2019R1A2C1003486 (D.-S.L.),), and a KIAS Individual Grant (No. CG079901) at Korea Institute for Advanced Study (D.-S.L.). This work is supported by the Center for Advanced Computation at Korea Institute for Advanced Study.
\end{acknowledgments}

\appendix

\section{Static-model networks with $\gamma$=3}
\label{app:static}

\begin{figure*}
\includegraphics[width=\textwidth]{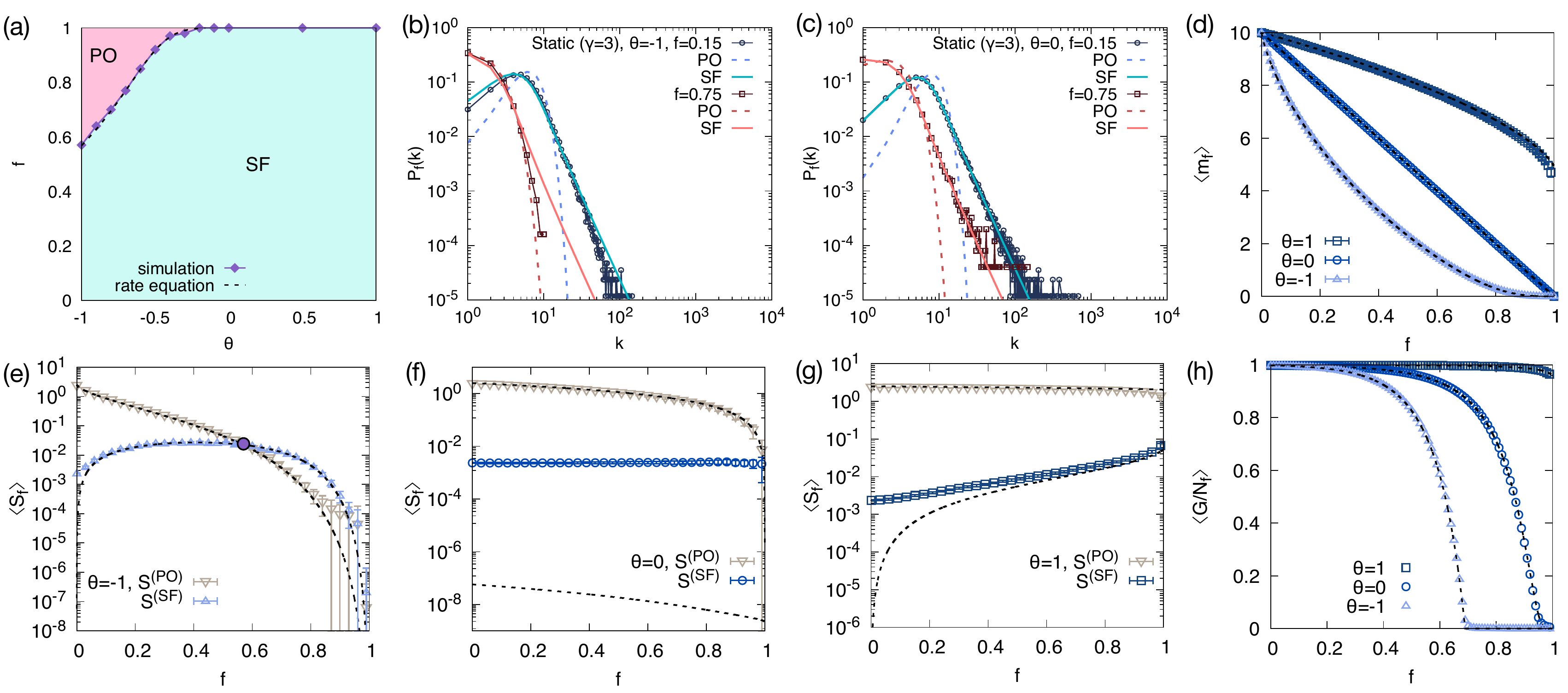}
\caption{Results for the original static-model SF networks~\cite{Goh2001} with the degree exponent $\gamma=3$.  Shown are (a) the regime diagram, (b, c) the degree distributions of the remaining subnetworks  for $\theta=-1$ and $\theta=0$, (d) the mean degree $\langle m_f \rangle$ as a function of the fraction $f$ of removed nodes, (e -- g) the relative entropies $\langle\spo\rangle$ and $\langle\ssf\rangle$ as functions of $f$, and (h) the relative size of the GCC, $\langle {G / N_f} \rangle$. 
}
\label{fig:app_gamma3}
\end{figure*}

We perform the same analyses as in the main text for the removal of nodes of the original static-model  networks~\cite{Goh2001} with $\gamma=3$ (the same exponent as the BA model), to see whether different network models can make a difference in the obtained results.   We present the results in Fig.~\ref{fig:app_gamma3}, which demonstrate that the main results are not changed. But some measures show a difference. The measures affected mainly by the strength of the hub nodes (the tail part of the degree distribution), such as the regime diagram, the shape of the degree distribution, the mean degree, and the relative size of GCC, turn out to be similar between the static model and the BA model sharing the same original degree exponent. The same is true for the relative entropy with respect to the PO distribution.  However, the functional behaviors of the relative entropy with respect to the SF distribution exhibit differences between the two model networks even if they share the same degree exponent.

To be specific, in Fig.~\ref{fig:app_gamma3}(a), the boundary between the PO and the SF regime for the static model is almost overlapped with that obtained for the BA networks in Fig.~\ref{fig:diagram}(b). The variations of the degree distribution and the mean degree with $f$ for the static model are also quite similar to the BA model results [compare Figs.~\ref{fig:app_gamma3}(b), \ref{fig:app_gamma3}(c), and \ref{fig:app_gamma3}(d) to Figs.~\ref{fig:degreedist}(a), \ref{fig:degreedist}(b), and \ref{fig:meank}(b), respectively]. Also, the relative size of the GCC is similar to the case of the BA model [compare Fig.~\ref{fig:app_gamma3}(h) and Fig.~\ref{fig:app_gc}(b)], which makes sense since the degree exponent $\gamma$ is the key control parameter for the percolation problem.

When it comes to the relative entropies of the degree distribution of the remaining subnetworks with respect to the two reference distributions, their behavior is not always similar between both original networks.  The  relative entropy $\spo$ with respect to the PO distribution decreases with $f$  almost in the same manner between  the static model and the BA model. However, the behaviors of the relative entropy $\ssf$ with respect to the SF distribution  are different between them.  Rather, the static-model networks with $\gamma=3$ [Figs.~\ref{fig:app_gamma3}(e), ~\ref{fig:app_gamma3}(f), and \ref{fig:app_gamma3}(g)] and $\gamma=2.5$ [Figs.~\ref{fig:entropy}(a), ~\ref{fig:entropy}(b), and ~\ref{fig:entropy}(c)] share similar behaviors of $\ssf$. When $\theta=1$ (hub-protecting removal), $\ssf$  increases with $f$ for the static-model networks with $\gamma=3$ while it decreases with $f$ for the BA networks. Compare Fig.~\ref{fig:app_gamma3}(g) and Fig.~\ref{fig:entropy}(f).  When $\theta=0$ (random removal), $\ssf$ is almost flat for the static model but it decreases with $f$ for the BA networks, as compared  between Fig.~\ref{fig:app_gamma3}(f) and Fig.~\ref{fig:entropy}(e). Also  in case of $\theta=-1$, the behavior of $\ssf$ for the static model with $\gamma=3$ [Fig.~\ref{fig:app_gamma3}(e)] is more similar to that for the static model with $\gamma=2.5$ [Fig.~\ref{fig:entropy}(a)] than to the BA model with $\gamma=3$ [Fig.~\ref{fig:entropy}(d)]. These discrepancies of the behaviors of the $\ssf$ between the static model and the BA model suggest that the alteration of the degree distribution during node removal, characterized by the relative entropy, depends on various structural properties including the degree exponent and the correlation of adjacent nodes' degrees.  

\section{Rate equation of degree distributions for general node removal processes}
\label{app:equation}

Here we present the rate equation of the degree distribution~\cite{Tishby2019, Tishby2020, KimJH2020, KimJH2022} for the generalized node removal process, the numerical solutions to which provide the approximated regime diagram and relative entropies in the thermodynamic limit; thus they complement the simulation results obtained for a finite system size. 

Let us consider a network composed of $N$ nodes and $L= N m_0/2$ links, represented by the adjacency matrix elements $a_{ij}$. At every discrete time step $\tau = 0,1,2, \ldots$, a  given node of degree $k$ is selected with probability  
\begin{equation}
q_k (\tau) = {(k+1)^{-\theta} \over \sum_{i\in I_{\rm surv} (\tau)} [k_i(\tau) +1]^{-\theta}}
\label{eq:removal_prob_f}
\end{equation}
and removed, where $I_{\rm surv}(\tau)$ is the set of surviving nodes at $\tau$ and $k_i(\tau)=\sum_{j\in I_{\rm surv}(\tau)} a_{ij}$ is the degree of node $i$ that only counts its surviving neighbors at $\tau$. Note that $q_k(\tau)$ in Eq.~(\ref{eq:removal_prob_f}) is the same as $q_f(k)$ in the main text with $f$ and $\tau$ related by $f=\tau/N$. Going from $\tau$ to $\tau+1$,  the number $N_k(\tau) \equiv \sum_{i\in I_{\rm surv}(\tau)} \delta_{k_i(\tau),k}$ of the surviving nodes of given degree $k$ will decrease by one if a node of degree $k$ is removed. Also,  as the neighbors of a removed node commonly lose a link and their degrees decrease by  one, $N_k(\tau)$ will increase (decrease) by the expected number of the nodes with degree $k+1$ (degree $k$) adjacent to the removed node, respectively. Therefore, $N_k(\tau)$ varies with time step $\tau$ as  
\begin{align}
N_k &(\tau+1) - N_k(\tau) = - N_k(\tau) q_k(\tau)  \nonumber\\
&+ (k+1) N_{k+1}(\tau) \tilde{q}_{k+1}(\tau) - k N_k(\tau) \tilde{q}_k (\tau) \,,
\label{eq:Nkdiff}
\end{align}
where we introduce the probability that a neighboring node of a node of degree $k$ is selected and removed at $\tau$ 
\begin{equation}
\tilde{q}_k(\tau) = {\sum_{k'} N_{kk'}(\tau) q_{k'}(\tau) \over \sum_{k'} N_{kk'}(\tau)} \,,
\end{equation}
where $N_{kk'}(\tau)$ is the number of links connecting nodes of degree $k$ and $k'$ at time $\tau$ (counted twice if $k\neq k'$)
\begin{equation}
N_{kk'}(\tau) \equiv \sum_{i,j\in I_{\rm surv}(\tau)} a_{ij} \delta_{k_i(\tau),k} \delta_{k_j(\tau),k'} \,.
\end{equation}

To consider the limit $N\to\infty$, we introduce a time-like continuous variable $f=\lim_{N\to\infty}{\tau / N}$  indicating the fraction of removed nodes and  define the fraction of the surviving nodes of degree $k$
\begin{equation}
D_f(k)\equiv \lim_{N\to\infty} {N_k(\tau = fN)\over N},
\label{eq:Dk}
\end{equation}
and the fraction of links connecting nodes with degrees $k$ and $k'$
\begin{equation}
D_f(k,k') \equiv \lim_{N\to\infty} {N_{kk'}(\tau=fN) \over 2L}.
\end{equation}
Note that these two functions are not summed  to one but to the total fraction of the surviving nodes $1-f$ and the total fraction of the links connecting the surviving nodes, respectively. Once the equation for $D_f(k)$ that we will present below is solved,  the degree distribution in the main text $P_f(k) = \lim_{N\to\infty} \left[ {N_k(\tau=fN) / \sum_{k'} N_{k'}(\tau=fN)} \right]$ can be obtained by  using the relation 
\begin{equation}
P_f (k) = {D_f(k)\over \sum_{k'} D_f(k')} = {D_f(k)\over 1-f}.
\label{eq:PkDk}
\end{equation}  

The equation for $D_f(k)$ is obtained by replacing $N_k(\tau)$ by $N D_f(k)$ and using the expansion $N_k(\tau+1) = N D_{f+df}(k) \simeq N D_f(k) + \partial D_f(k) / \partial f$ with $df = N^{-1}$ in Eq.~\eqref{eq:Nkdiff}, which is given by
\begin{align}
{\partial \over \partial f} & D_f(k) = - D_f (k) r_f(k)  \nonumber\\
&+ (k+1) D_f(k+1) \tilde{r}_f (k+1) - k D_f(k) \tilde{r}_f (k)
\label{eq:Dkdf}
\end{align}
in a similar form to Eq.~\eqref{eq:Nkdiff} with 
\begin{align}
r_f(k) \equiv \lim_{N\to\infty} N q_k(\tau=fN) &={(k+1)^{-\theta} \over \sum_{k'} D_f (k') (k'+1)^{-\theta}} \nonumber \,, \\
\tilde{r}_f (k) \equiv \lim_{N\to\infty} N \tilde{q}_k(\tau=fN)&= {\sum_{k'} D_f (k,k')\, r_f(k') \over \sum_{k'} D_f (k,k')} \,,
\label{eq:rk}
\end{align}
meaning the rate that a node of degree $k$ is removed and the rate that a neighbor of a node of degree $k$ is removed, respectively. 

A problem in solving Eq.~\eqref{eq:Nkdiff} or Eq.~\eqref{eq:Dkdf} for $N_k(\tau)$ or $D_f(k)$ is that it depends on an unknown two-point function $N_{kk'}(\tau)$ or $D_f(k,k')$ through $\tilde{q}_k(\tau)$ or $\tilde{r}_f(k)$, which in turn depends upon a three-point function, etc. To truncate the hierarchy,  we take the approximation that 
\begin{equation}
N_{kk'}(\tau) = {k k' N_k(\tau) N_{k'}(\tau) \over 2L(\tau) },
\label{eq:Nkkapprox}
\end{equation}
where $L(\tau) =(1/2)\sum_{k} kN_k(\tau)$ is the number of links connecting the surviving nodes, or equivalently 
\begin{equation}
D_f(k,k') = {kk' D_f(k) D_f(k') \over (1-f) \,  m_0\, m_f },
\label{eq:Dkkapprox}
\end{equation}
with $m_f = \sum_k k P_f(k) = (1-f)^{-1} \sum_k k D_f(k)$. These approximations are valid when the degrees of adjacent nodes are not correlated, being independent of each other.  Inserting Eq.~\eqref{eq:Dkkapprox} into Eq.~\eqref{eq:rk}, we find that $\tilde{r}_f(k)$ is  represented in terms of $D_f(k)$ as 
\begin{equation}
\tilde{r}_f(k) \equiv {\sum_{k'} k' \, D_f(k') \, (k'+1)^{-\theta} \over \sum_{k'} k' D_f (k') \sum_{k''} D_f (k'') (k''+1)^{-\theta} \, } = \tilde{r}_f,
\label{eq:rkapprox}
\end{equation}
which is independent of $k$.

For $D_f(k)$, we numerically solve Eq.~\eqref{eq:Dkdf} with $r_f(k)$ given in Eq.~\eqref{eq:rk} and $\tilde{r}_f$ given in Eq.~\eqref{eq:rkapprox}. Choosing a sufficiently small value of $df = 10^{-7}$  and a large value of $k_{\rm max}=10^5$, we compute recursively $D_{f+df}(k) = D_f(k) +  df \partial D_f(k) / \partial f$ from $D_f(k)$ for all $k\leq k_{\rm max}$ with the boundary condition $D_f(k_{\rm max}+1)=0$~\cite{KimJH2020}. The degree distribution $P_f(k)$ is then obtained from $D_f(k)$ by Eq.~\eqref{eq:PkDk}, i.e., $P_f(k) = D_f(k)/\sum_{k'=0}^{k_{\rm max}} D_f(k')$.  Note that $P_f(k)$ indicates the fraction of nodes of degree $k$ among the surviving nodes and satisfies $\sum_{k=0}^{k_{\rm max}} P_f(k)=1$. 

With the obtained $P_f(k)$, we compute the mean degrees [Fig.~\ref{fig:meank}] and the relative entropies [Fig.~\ref{fig:entropy}] as functions of $f$, and obtain the regime diagrams from the results [Fig.~\ref{fig:diagram}]. These numerical solutions show a good agreement with the simulation results when the original networks are the static-model SF networks, implying that the degree correlations are indeed negligible. On the other hand, there is some deviation between the numerical and simulation results for the BA networks, probably due to their significant degree correlations between the adjacent nodes. 

Another caveat of the numerical solutions is that due to the large but finite value of $k_{\rm max}=10^{5}$, the measured value of $m_f = \sum_{k=0}^{k_{\rm max}} k P_f(k)$ is smaller than the value that would be measured with $k_{\rm max}\to\infty$. Its remarkable consequence is that even an original SF network does not have a zero relative entropy $S^{\rm (SF)}$  with respect to the SF distribution at $f=0$ [Figs.~\ref{fig:entropy}(a)--\ref{fig:entropy}(c)]; an initial network with the degree distribution $P_0(k) = P^{\rm (SF)}(k;\gamma;m_0^{\rm (orig)})$ is compared with the reference SF reference distribution $P^{\rm (SF)}(k;\gamma;m_0\lesssim m_0^{\rm (orig)})$ of a slightly smaller mean degree with the difference estimated as $m_0^{\rm (orig)} - m_0  = \sum_{k=k_{\rm max}+1}^\infty k P^{\rm (SF)} (k;\gamma;m_0^{\rm (orig)})\sim k_{\rm max}^{2-\gamma}$. Furthermore, we note that due to the limitation of computing resource and time, we compute the SF distribution according to Eq.~\eqref{eq:randomsf} for $k\leq 10^3$ and adopt the asymptotic behaviors $P^{\rm (SF)} (k;\gamma;m_f) \simeq (\gamma-1) \left(m_f {\gamma-2\over \gamma-1}\right)^{\gamma-1} k^{-\gamma}$ for $k>10^3$ as given in Ref.~\cite{Lee2004}. For the original BA networks with mean degree $2n$, we take their exact degree distribution  $P^{\rm (BA)}(k; 2n)=2n(n+1)/k(k+1)(k+2)$~\cite{Dorogovtsev2000} for all $k_{\rm min} = n \leq k \leq k_{\rm max}=10^5$.  Finally, we calculate the size of the giant component [Fig.~\ref{fig:app_gc}] using a formal generating function method~\cite{PhysRevLett.85.5468, Newman2001}.

\end{CJK}

\end{document}